\documentclass[12pt]{article}
\usepackage{jhep-mod}  

\usepackage{bm}   
\usepackage{amssymb,amsmath,amsthm}  
\usepackage{mathrsfs}  
\usepackage[utf8]{inputenc}
\usepackage{enumerate}
\usepackage{appendix} 
\usepackage{graphicx}
\usepackage{dcolumn} 
\usepackage{bm}
\usepackage{multirow}
\usepackage{float}
\usepackage{tikz}
\usepackage{setspace}
\usepackage[normalem]{ulem} 
\usepackage{tocloft}
\definecolor{purple}{rgb}{1,0,1}
\definecolor{lime}{HTML}{A6CE39} 

\newcommand{\blue}[1]{{\color{blue} #1}}

\definecolor{lime}{HTML}{A6CE39}
\newcommand{\orcidicon}{%
	\begin{tikzpicture}
	\draw[lime, fill=lime] (0,0) 
		circle [radius=0.16] 
		node[white] {{\fontfamily{qag}\selectfont \tiny ID}};
	\draw[white, fill=white] (-0.0625,0.095) 
		circle [radius=0.007];
	\end{tikzpicture}
	\hspace{-5mm}
}
\newcommand\orcidPrado{{\href{https://orcid.org/0000-0001-8073-4896}{\orcidicon}}}
\newcommand\orcidMatt{{\href{https://orcid.org/0000-0003-1088-6485}{\orcidicon}}}
\begin{document}
\title{
\leftline{\huge Hawking--Ellis classification of stress-energy:}
{\huge test-fields versus back-reaction}\\
}

\author{\Large Prado Mart\'{\i}n-Moruno$^1$\orcidPrado{}{\sf and} 
Matt Visser$^2$\orcidMatt{}}
\affiliation{$^1$ Departamento de F\'isica Te\'orica and IPARCOS,  Universidad Complutense de Madrid, \\ \null\quad E-28040 Madrid, Spain}
\emailAdd{pradomm@ucm.es}
\affiliation{$^2$ School of Mathematics and Statistics, Victoria University of Wellington, \\ \null\quad PO Box 600, Wellington 6140, New Zealand}
\emailAdd{matt.visser@sms.vuw.ac.nz}
\abstract{
\parindent0pt
\parskip7pt
\null\\
We consider the  Hawking--Ellis (Segr\'e--Pleba\'nski) classification of stress-energy tensors, both in the test-field limit, and in the presence of back-reaction governed by the usual Einstein equations. 
For test fields it is not too difficult to get a type~IV stress-energy via quantum vacuum polarization effects. (For example, consider the Unruh quantum vacuum state for a massless scalar field in the Schwarzschild background.)
However, in the presence of back-reaction driven by the ordinary Einstein equations the situation is often much more constrained. 
For instance: (1) in any static spacetime the stress-energy is always type I in the domain of outer communication, and on any horizon that might be present; (2) in any stationary axisymmetric spacetime the stress-energy is always type I on any horizon that might be present; (3)  on any Killing horizon that is extendable to a bifurcation 2-surface the stress-energy is always type I;
(4) in any stationary axisymmetric spacetime the stress-energy is always type I on the axis of symmetry; (5) \emph{some} of the homogeneous Bianchi cosmologies are guaranteed to be Hawking--Ellis type I (for example, all the Bianchi type I cosmologies, all the FLRW cosmologies, and all the ``single mode'' Bianchi cosmologies). 
That is, in very many physically interesting situations once one includes back-reaction the more unusual stress-energy types are automatically excluded. 

\medskip
{\sc Date:}  Saturday 27 February 2021; Friday 5 March; \LaTeX-ed \today

test-fields; back-reaction.

}
\maketitle
\def\d{{\mathrm{d}}}
\def\tr{{\mathrm{tr}}}
\def\O{{\mathcal{O}}}
\parindent0pt
\parskip7pt
\clearpage
\section{Introduction}

The energy conditions are inequalities on the stress-energy tensor that are requested to be satisfied by orthodox material content~\cite{LNP-survey}. Although they are not derived from fundamental physics~\cite{twilight}, they can be useful to extract general consequences.  In particular, in the framework of a theory of gravity, they allow us to restrict attention to the physically reasonable spacetimes and their characteristics~\cite{book}. Nowadays one of the more conservative energy conditions, the strong energy condition (SEC), is known to be violated macroscopically during cosmic evolution~\cite{Visser:1997, Visser:1997-b, Visser:1999, Cattoen:2006}. Furthermore it is well-known that the weakest energy condition, the null energy condition (NEC), can be violated by the renormalized stress-energy tensor when considering quantum effects \cite{Martin-Moruno:semiclassical}. However, it seems that if semi-classical physics could allow unrestricted violations of the null energy condition, one should appeal to additional physics to forbid potential problems of the spacetime, such as causal paradoxes.

Historically, almost all numerical and semi-analytic computations of the renormalized stress-energy tensor have been preformed in the test-field approximation; \emph{ie}, picking a fixed background and ignoring back-reaction. See for example references~\cite{book} and~\cite{ Visser:1996-I, Visser:1996-II, Visser:1996-III, Visser:1997-IV, Visser:1997-V}. 
Working in the test field approximation it is relatively straightforward to see that all the classical energy conditions can be violated, and more subtly that one can encounter the more \emph{outr\'e} of the stress-energy tensors occurring in the Hawking--Ellis (Segr\'e--Pleba\'nski) classification \cite{H&E}.  In particular, as discussed below, one can encounter type~IV stress-energy tensors \cite{Martin-Moruno:semiclassical, Roman86,Don-Page:2016}, which necessarily violate all the energy conditions \cite{LNP-survey}.

In counterpoint, recently Hideki Maeda realised that once one includes back-reaction and looks at self-consistent solutions to the Einstein equations the situation simplifies tremendously~\cite{Maeda:2020}.
Specifically, Hideki Maeda showed that in any static spacetime solving the Einstein equations, then in those regions where the $t$ coordinate is actually timelike,  the stress-energy is always of type~I according to the Hawking--Ellis (Segr\'e--Pleba\'nski) classification.
That is: The stress-energy tensor is certainly of type~I in the so-called ``domain of outer communication'', but one has to be careful below any horizon that might be present.
Therefore, in the region of interest, for static spacetimes the stress-energy tensor (of either classical or semi-classical fields) takes the simplest form possible; and, therefore, there is no reason \emph{a priori} why one should think that the weakest energy conditions would necessarily have to be violated.

\clearpage
Herein we shall generalize this result in various ways:
\begin{itemize}
\itemsep-3pt
\item 
 In any static spacetime the stress-energy is always type I in the domain of outer communication, and on any horizon that might be present.
 \item 
 In any stationary spacetime the stress-energy is always type I on any horizon that might be present.
\item  
The stress-energy is always type I on any bifurcate Killing horizon.
 \item
 In any stationary axisymmetric spacetime the stress-energy is always type I on the axis of symmetry.
 \item
 Some of the homogeneous Bianchi cosmologies are guaranteed to be Hawking--Ellis type I (for example, all the Bianchi type I cosmologies, all the FLRW cosmologies, and all the ``single-mode'' Bianchi cosmologies). 
Other Bianchi cosmologies require a case-by-case analysis.
\end{itemize}
That is, in very many physically interesting situations the more \emph{outr\'e} stress-energy types are automatically excluded. 
While we shall largely focus on (3+1) dimensions, some brief comments on (2+1) and (1+1) dimensions are relegated to the appendix.
Finally, note that many proposed modifications or extensions of Einstein gravity have equations of motion that 
can be rearranged into the form $G_{ab} = 8 \pi [T_\mathrm{effective}]_{ab}$.
For example, consider Maeda's $f(g_{ab},R_{abcd})$ gravity~\cite{Maeda:2020}, the more specific $f(R)$ theories of gravity~\cite{Sotiriou:2008}, Rastall gravity~\cite{Rastall:1972,Rastall:1976,Rastall:1977,Rastall=Einstein}, Horndeski theories~\cite{Horndeski:1974,McManus:2016},  and the like~\cite{Gleyzes:2014}.
Then at a minimum our results imply that  $[T_\mathrm{effective}]_{ab}$ is Hawking--Ellis 
type I under the stated conditions.
(And to say anything more about $T_{ab}$ itself, we would need to specify 
the precise extended/modified gravity model in question.)

\section{Test field stress-energy tensors}\label{S:test}

The use of test fields (and even test particles) in general relativity has a very long and distinguished history. Even the seemingly utterly innocuous statement that falling particles follow spacetime geodesics is ultimately limited to the test-particle regime~\cite{Infeld}.  In a quantum field theoretic context, Hawking's original derivation of Hawking radiation is intrinsically a test-field calculation~\cite{Hawking-explosions, Hawking-radiation}. 

\subsection{Energy conditions}\label{SS:energy}

The classical energy conditions of general relativity are used to keep some of the more \emph{outr\'e} aspects of gravitational physics somewhat under control.
While the classical energy conditions are not truly fundamental physics~\cite{twilight},
they are nevertheless very useful to work with~\cite{Visser:1997,twilight,book,Visser:1997-b,Visser:1999, Cattoen:2006, Fewster:2002, Roman:2004, LNP-survey, Martin-Moruno:semiclassical, Martin-Moruno:2013sfa, Martin-Moruno:2015,Curiel:2014}.

Cosmologically, to make the ``accelerating universe'' compatible with general relativity one has to violate the strong energy condition (SEC) on cosmological scales~\cite{Visser:1997,Visser:1997-b,Visser:1999}. This is not surprising once one notes that SEC is the requirement of gravity always  being attractive in general relativity, which is obviously not the case during inflation and the current cosmic epoch. Its mathematical expression is
\begin{equation}\label{NEC}
V_a\left(T^a{}_b-\frac{1}{2}T\delta^a{}_b\right)V^b\geq0,
\end{equation}
for any timelike vector $V^a$. 

On the other hand,
we emphasise that to make the existence of Hawking radiation compatible with the black hole area increase theorem, one must violate the null energy condition (NEC) through quantum effects. The NEC is the weakest of the point-wise energy conditions. It consists of demanding 
\begin{equation}\label{NEC}
k_a\,T^a{}_b\,k^b\geq0,
\end{equation}
for any null vector $k^a$.
It is a limiting case of the weak energy condition (WEC), which states that the energy density measured by any observer has to be non-negative ($V_a\,T^a{}_b\,V^b\geq0$), and of SEC.
In view of these issues, while this was not at first fully appreciated, it is just as well that explicit test-field numerical calculations of the renormalized stress-energy tensor in Schwarzschild and related backgrounds straightforwardly lead to numerical violations of all of the classical point-wise energy conditions~\cite{book, Visser:1996-I, Visser:1996-II, Visser:1996-III, Visser:1997-IV, Visser:1997-V}. (We recommend the reader interested in modern investigations on newer variants of the  energy conditions to check references~\cite{Martin-Moruno:semiclassical} and~\cite{Martin-Moruno:Rainich,Martin-Moruno:2013sfa,Maeda:2018}.)

\subsection{Hawking--Ellis (Segr\'e--Pleba\'nski) classification}\label{SS:hell}

The Hawking--Ellis (Segr\'e--Pleba\'nski) classification of stress-energy tensors is composed of four different types \cite{H&E}. Each type represents a different eigenvector structure. Those are: type~I, which has 1 timelike and 3 spacelike eigenvectors; type~II, with 1 double null eigenvector and 2 spacelike eigenvectors; type~III, which has 1 triple null eigenvector and 1 spacelike eigenvector; and type~IV, which has no causal eigenvector. Using Lorentz transformations, all these types can (in an orthonormal basis) be put into the form (see, for example, reference~\cite{LNP-survey})
\begin{equation}\label{SET-HE}
T_{\hat a\hat b} = \left[\begin{array}{ccc|c} 
\rho & f_1 & f_2 &0 \\ f_1 &p_1 &\sigma &0\\ 
f_2&\sigma& p_2 & 0\\ \hline
0&0&0&p_3
\end{array}\right].
\end{equation}
Here $f_1=f_2=\sigma=0$ for type~I; $f_1\neq0$, $f_2=\sigma=0$, and $\rho+p_1=2f_1$ for type~II; $f_1=0$, $f_2\neq0$ and $\sigma\neq0$ for type~III; and  $f_1\neq0$, $f_2=\sigma=0$, and $p_1=-\rho$ for type~IV.

Investigating which of these Hawking--Ellis types a given stress-energy tensor belongs to can be somewhat tricky for the renormalized stress energy tensors obtained by numerical methods, at least partly due to delicate numerical artefacts, and the potential for delicate numerical (almost) cancelations. 

Working in spherical symmetry for now let us identify the orthonormal components of the stress-energy as 
\begin{equation}\label{SET-SS}
T_{\hat a\hat b} = \left[\begin{array}{cc|cc} 
\rho & f & 0 &0 \\ f &p_\parallel &0 &0\\ \hline
0&0& p_\perp & 0\\
0&0&0&p_\perp
\end{array}\right].
\end{equation}
So, type~III is already discarded in spherically symmetric situations.
The Lorentz invariant eigenvalues, defined by 
\begin{equation}
\det(T_{\hat a\hat b} - \lambda g_{\hat a\hat b} )=0,
\end{equation}
are easily calculated to be
\begin{equation}
\lambda \in \left\{ \frac{1}{2}(p_\parallel-\rho) \pm \frac{1}{2}\sqrt{(\rho+p_\parallel)^2-4 f^2};
\;\; p_\perp,\;  p_\perp \right\}.
\end{equation}
The \emph{sign} of the quantity 
\begin{equation}
\Gamma = (\rho+p_\parallel)^2-4 f^2
\end{equation}
is thus key to controlling the reality and degeneracy of the eigenvalues, which in turn controls the Hawking--Ellis type of the stress-energy tensor. 

\subsubsection{Hawking--Ellis type~I}\label{SS:I}

For a type~I stress-energy tensor there is always some observer who sees that there are no net fluxes, so such a stress-energy tensor can be fully diagonalized via Lorentz transformations. Most classical fields that can be found in nature have a type~I stress-energy tensor, and for type~I stress-energy tensors one can always find an observer ``moving with the field''. Since in this situation $f=0$, we have $\Gamma>0$ leading to 2 different eigenvalues in addition to $p_2$ and $p_3$. (One also has the degenerate case in which $\rho=-p_\parallel$, that still has four eigenvectors.)

On the other hand, the renormalized stress-energy tensors of quantum vacuum states for which the flux term can be made to vanish by Lorentz transformation are also type~I.
Since there is no net flux in either the Hartle--Hawking~\cite{Visser:1996-I} or Boulware~\cite{Visser:1996-II} quantum vacuum states, the relevant stress-energy is automatically diagonal, and so of Hawking--Ellis type I. 

\subsubsection{Hawking--Ellis type~II}\label{SS:II}

For type~II stress-energy tensor we cannot find any timelike observer for which the flux term vanishes.
If $\rho + p_\parallel  =\pm  2   f$, then $\Gamma=0$ and the Lorentz invariant eigenvalues of the stress-energy are
\begin{equation}
\lambda \in \left\{  \frac{1}{2}(p_\parallel-\rho), \;  \frac{1}{2}(p_\parallel-\rho)   ;\;\;  p_\perp,\;  p_\perp \right\}.
\end{equation}
This implies the stress-energy would be Hawking--Ellis type II.  Classical radiation and other zero-mass fields have type~II stress-energy tensors. 

Regarding the test-field renormalized stress-energy tensors, unfortunately this situation, even if it were to arise in theory, would in practice be unstable to numerical roundoff error \cite{Martin-Moruno:essential}.
(Any roundoff error in the stress-energy components would generically lift the degeneracy in the eigenvalues --- unless one has a symmetry principle keeping the equality of the eigenvalues exact.)
In short, numerically computed renormalized stress-energy tensors are not a useful diagnostic for exploring Hawking--Ellis type~II stress-energy tensors.

\subsubsection{Hawking--Ellis type~III}\label{SS:III}

For this type of stress-energy tensor one cannot find any timelike observer that measures no fluxes and no stresses. Therefore, spherical symmetry is not compatible with a type~III stress-energy tensor.
In classical physics there are no natural matter fields that lead to this type of tensors, although artificial models have been constructed \cite{Martin-Moruno:2019kzc}.  

\clearpage
For the renormalized stress-energy tensors,
to open the door for type~III stress-energy tensors, we should go beyond spherically symmetric quantum vacuum states; for example, by considering a Kerr background. 
In this case, one has a renormalized stress-energy tensor (for a minimally coupled massless scalar field) for the Kerr geometry \cite{Levi:2016exv} that can in principle be of type~III.
It should be noted, however, that even if not explicitly forbidden by symmetries, type~III is (like type~II) unstable to numerical roundoff error \cite{Martin-Moruno:essential}. 
Numerically computed renormalized stress-energy tensors are not a useful diagnostic for exploring Hawking--Ellis type~III stress-energy tensors.

\subsubsection{Hawking--Ellis type~IV}\label{SS:IV}

In this case $\Gamma<0$, so there are no causal eigenvectors. There are 2 real spacelike and 2 complex eigenvectors, therefore, no causal (timelike or null) eigenvectors.
Type IV stress-energy tensors can be understood as a complex extension of type I, since in order to get an observer measuring zero fluxes, she/he would need be able to measure complex energy densities and pressures. Therefore, there is no physical timelike observer who measures no fluxes.

There are no known classical matter fields that lead to a type~IV stress-energy tensor. Regarding renormalized stress-energy tensors of quantum vacuum state, as already mentioned, all cases without an intrinsic flux are automatically type~I. So, this leaves us with only the Unruh~\cite{Visser:1997-IV}  quantum vacuum state to investigate --- and \emph{a priori} the relevant stress-energy could be any one of types I, II, or IV.

Up to the best of our knowledge, the first investigation about a type~IV stress-energy tensor was made by Roman in 1986 \cite{Roman86}. Establishing an analogy with a model describing a singularity-free collapsing star, he argued that the renormalized vacuum expectation value of spherically symmetric evaporating black hole has to be type~IV in a neighborhood of the apparent horizon (see section II of reference \cite{Roman86}). 
Unaware of that result, in reference \cite{Martin-Moruno:semiclassical} we proved that the stress-energy tensor of the Unruh vacuum is type~IV far from the horizon. However, we found a spurious pole in $\Gamma$ due to round-off errors in the numerical calculations.
Later on, in reference \cite{Don-Page:2016}, Abdolrahimi, Page, and Tzounis showed that the stress-energy tensor for the massless conformal scalar in the Unruh state is Type IV everywhere outside the horizon.

\enlargethispage{10pt}
\subsection{Summary: test field stress-energy tensors}\label{SS:test field summary}
In short, in the test-field framework Hawking--Ellis types I and IV are quite common. 
In contrast, while Hawking--Ellis types II and III are not absolutely forbidden in the test-field framework, they are numerically and perturbatively unstable. 

\section{Back-reaction via the Einstein equations}\label{S:back-reaction}

Once one introduces back-reaction by imposing the Einstein equations
\begin{equation}
T_{ab} = {1\over 8\pi G_N} \; G_{ab},
\end{equation}
the stress-energy becomes very tightly constrained in terms of the spacetime geometry. 
Purely geometrical considerations will then fully control the energy conditions and the Hawking--Ellis classification.

\subsection{Spacelike slicings}
Let us seek to understand what it means geometrically to have a type~I stress-energy tensor. Supposing that we have a spacelike slicing of spacetime, we can adopt the ADM formalism and write
\begin{equation}
\d s^2 = - N^2 \d t^2+2 N_i\, \d t \,\d x^i +  g_{ij}\, \d x^i \d x^j.
\end{equation}
Here $N$ is the lapse function, $N^i$ is the shift vector, and $g_{ij}$ is the intrinsic 3-geometry.
The normal to the spacelike slices is the timelike co-vector $n_a = (\nabla t)_a /|\nabla t|$, the projection operator onto the 3-hypersurfaces is
\begin{equation}
h^a{}_b=g^a{}_{b}+n^a n_b,
\end{equation}
and the extrinsic 3-geometry curvature is 
\begin{equation}\label{Kij}
K_{ij}=h^a{}_ih^b{}_j \,n_{a;b},
\end{equation}
where $;$ denotes (3+1) dimensional covariant derivative. One key result of the ADM formalism is that
$n_a R^{ab} h_{bc}$
can be evaluated in terms of intrinsic derivatives of the extrinsic curvature $K_{ij}$. From the Gauss--Coddazzi equations, one has
\begin{equation}\label{Rni}
R^{ni} =  K^{ij}{}_{:j} -  g^{ij} K_{:j},
\end{equation}
where we have denoted $R^{ni}=n_aR^{ab}h^i{}_b$ and now $:$ denotes covariant derivative in the 3-space of the spacelike slices. As one has
\begin{equation}
G_{ab}\,n^a h^b_c= \left(R_{ab}-\frac{1}{2}Rg_{ab}\right)n^a h^b{}_c=R_{ab}n^a h^b{}_c-\frac{1}{2}Rg_{ab}n^a (g^b{}_c+n^bn_c)=R_{ab}\,n^a h^b{}_c,
\end{equation}
it follows that
\begin{equation}\label{Gni}
G^{ni} =  K^{ij}{}_{:j} -  g^{ij} K_{:j}.
\end{equation}

Certainly $G^{ni}=0$ is a \emph{sufficient} condition to guarantee the stress-energy is type~I. 
All such geometries have a stress-energy tensor of the form 
$R_{ti}=0=G_{ti}$, so that
\begin{equation}
T_{ab} = \left[ \begin{array}{c|c} 
\rho & 0 \\ \hline 0 & T_{ij} 
\end{array} \right].
\end{equation}
Using ordinary spatial 3-rotations, this is now manifestly Hawking--Ellis type~I. 

In reference \cite{Maeda:2020} Hideki Maeda uses hypersurface orthogonality to get the spatial slicing, and  then adds the Killing condition to enforce $K_{ij}\to 0$. 
However, it is clear from the discussion above that it is more than sufficient to demand the somewhat weaker condition
\begin{equation}\label{condition}
K^{ij}{}_{:j} -  g^{ij} K_{:j} =0.
\end{equation}
That is, one might reasonably expect to have many different classes of spacetimes that satisfy this sufficient condition. We shall explore various possibilities below.

\subsection{Static spherically symmetric spacetimes}\label{SS:sss}

This particular case is trivial. For a static spacetime there exists a spacelike hypersurface 
(3-surface) orthogonal to the timelike Killing vector. If one additionally requires spherical symmetry, then that 3-surface retains this symmetry. Then, one can always choose a coordinate system where the metric is diagonal and time independent. In that specific coordinate system the Einstein tensor (and so the stress-energy tensor) is also diagonal --- so the stress-energy is automatically Hawking--Ellis type I. The physically more interesting question is what happens when one relaxes these stringent conditions. 

\subsection{Static spacetimes --- domain of outer communication}\label{SS:static}

The existence of a hypersurface orthogonal Killing vector implies that, in the region where that Killing vector is timelike, one can choose coordinates $x^a=(t,x^i)$ to write
\begin{equation}
g_{ab} = \left[\begin{array}{c|c} - N^2 & 0 \\ \hline 0 & g_{ij} \end{array} \right],
\end{equation}
with $\dot N=0$ and $\dot g_{ij}=0$.
That is, the (3+1) metric is block diagonalizable into $(1\times1)\oplus(3\times3)$ blocks, where the 3-metric $g_{ij}$ has Euclidean signature, and the components of the metric in this decomposition are time independent.
A quick version of the argument leading to this conclusion is given in reference~\cite[page 119]{Wald}. 
In these coordinates the timelike Killing vector is $K^a = (1,0,0,0)^a$ with, by definition, $g_{ab} K^a K^b = - N^2$.
The Killing covector is $K_a = (-N^2;0,0,0)_a = -N^2 \nabla_a t$.

(Notice block diagonalizability of the metric is a \emph{choice}. For instance, in static spherical symmetry the metric is block-diagonal  in curvature coordinates,
conformal coordinates, isotropic coordinates, and proper distance coordinates; 
but the metric is \emph{not} block-diagonal in Painleve--Gullstrand and Eddington--Finklestein coordinates.)

In this situation block diagonalizability of the metric implies block diagonalizability of the Ricci tensor.
To most easily see this one appeals to the Gauss--Coddazzi equations as argued in the previous section. In the coordinate system set up above, the unit timelike vector normal to the spacelike hypersurfaces is $n^a=\frac{1}{N}K^a$ and, therefore, the extrinsic curvature $K_{ij}$ of the constant-$t$ 3-surfaces is zero,
\begin{equation}
K_{ij}=\frac{1}{2N}\;\dot g_{ij}=0.
\end{equation}
Specifically this means,
through equation (\ref{Rni}), $R_{ti}=0$, so that
\begin{equation}
R_{ab} = \left[\begin{array}{c|c} R_{tt}& 0 \\ \hline 0 & R_{ij} \end{array} \right].
\end{equation}
Consequently we also have
\begin{equation}
G_{ab} = \left[\begin{array}{c|c} G_{tt}& 0 \\ \hline 0 & G_{ij} \end{array} \right].
\end{equation}
Applying the Einstein equations, the stress-energy tensor satisfies
\begin{equation}
T_{ab} = \left[\begin{array}{c|c} T_{tt}& 0 \\ \hline 0 & T_{ij} \end{array} \right].
\end{equation}
The remaining $3\times3$ block $T_{ij} $ can be diagonalized using ordinary 3-space rotations, 
so the stress-energy is automatically type~I, (in the region where the $t$ coordinate is timelike). Because $T_{ab} $ has a timelike eigenvector types Hawking--Ellis II, III, and IV are explicitly excluded.
(Similarly, in the domain of outer communication of any static spacetime, where the Killing vector is timelike, all of the polynomial curvature invariants are guaranteed to be poitive semidefinite~\cite{novel}.)
Note that block diagonalizability \emph{and} time independence are both necessary for this particular proof.

On the other hand, below any horizon that might be present the $t$ coordinate is spacelike, and so 
the 4-metric (while it is still block diagonalizable) is then of the form
\begin{equation}
g_{ab} = \left[\begin{array}{c|c} + N^2 & 0 \\ \hline 0 & g_{ij} \end{array} \right].
\end{equation}
Here $g_{ij}$ is now a (2+1)-dimensional Lorentzian signature metric. 
The Ricci tensor and Einstein tensor are still block diagonalize, but now 
the Killing vector $K^a$ is spacelike, and so the stress-energy has at least one spacelike eigenvector --- and so it cannot be deduced that the stress-energy is type~I,  in fact below the horizon it could in principle be any one of the Hawking--Ellis types~I to~IV.

\subsection{On-horizon stress-energy}

Let us now investigate what happens to the stress-energy on any horizon that may be present. 
The key point is that on any horizon that might be present there is an enhanced symmetry and the stress-energy takes the restricted form
\begin{equation}
\label{restricted}
[T_{\hat a\hat b}]_H = \left[ \begin{array}{cc|c} 
\rho_H & 0 & 0 \\ 0 &-\rho_H & 0 \\ \hline 0 & 0 & [T_{\hat i\hat j}]_H
\end{array} \right],
\end{equation}
where the hats denote that the tensor is expressed in an orthonormal basis.
This is enough to guarantee the stress-energy is Hawking--Ellis type~I.

\subsubsection{Horizons in static spacetimes}

In the previous section we have seen that the stress-energy tensor of static spacetimes is type~I in the domain of outer communication, that is where the hypersurface orthogonal Killing vector is timelike. Now we will see what happens on the horizon,  where the hypersurface orthogonal Killing vector is null.
To see that the claimed result holds in spherical symmetry is essentially trivial~\cite{dirty}. 
Note that in this situation the metric can be written on the form
\begin{equation}
\d s^2 = - \exp\{+2\Phi(r)\}\;\left(1-{2m(r)\over r}\right) \d t^2 +{\d r^2\over1-2m(r)/r} + r^2 \d\Omega^2.
\end{equation}
A quick computation yields
\begin{equation}
\rho + p_\parallel = {\Phi'(r) (r-2m(r))\over4\pi r^2},
\end{equation}
which vanishes on the horizon at $r=2m(r)$. 

To see that this still works for arbitrary static spacetimes is considerably trickier~\cite{dirty2}.
(The current argument is similar but not quite identical to that in reference~\cite{dirty2}.)
First use the static condition to write $\d s^2 = - N(x^i) \d t^2 + g_{ij}\,\d x^i \d x^j$. 
Next, within the spatial 3-slices, use the level-sets of $N(x^i)$ to define one of the coordinates $z$,
the remaining 2 coordinates being $x^\mu=(x,y)$. That is our coordinates are $x^a=(t,x^i) = (t,z,x^\mu) = (t,z,x,y)$ and without loss of generality
\begin{equation}
\d s^2 = - f(z) \,\d t^2 +{\d z^2\over f(z)} + 2 N_\mu(x^k) \, \d z\, \d x^\mu + g_{\mu\nu}(x^k) \, \d x^\mu \d x^\nu.
\end{equation}
(Note that this is just a redefinition of $z$ to get the desired form for $g_{zz}$.)
We still have coordinate freedom in the $x^\mu=(x,y)$ 2-surfaces which we can use to 
set $N_\mu\to 0$. 
That is, since the metric is of the form $(1\times1)\oplus(3\times3)$, we can always diagonalize the $(3\times3)$ block that is a symmetric matrix in Euclidean space.
Thence
\begin{equation}
\d s^2 = - f(z) \,\d t^2 +{\d z^2\over f(z)}  + g_{\mu\nu}(x,y,z) \,\d x^\mu \d x^\nu.
\end{equation}
Now, let us go to an orthonormal non-coordinated basis $\{e^{\hat a}\}$. Since $g_{ab}=\eta_{\hat a \hat b}e^{\hat a}{}_ae^{\hat b}_b$, we have
\begin{equation}
T_{\hat a\hat b}=e_{\hat a}{}^a e_{\hat b}{}^b \,T_{ab} .
\end{equation}
A brief but slightly tedious computation now yields
\begin{equation}
T_{\hat a\hat b} = \left[ \begin{array}{c|c|cc} 
\rho & 0 & 0 & 0\\ \hline 0 &\;-\rho +\O(f) & \;\O(\sqrt{f}) &\;\O(\sqrt{f})\\ \hline 
0 & \O(\sqrt{f})  & \O(1) &\O(1) \\ 0 & \;\O(\sqrt{f}) & \O(1) &\O(1)
\end{array} \right],
\end{equation}
where $p_{\hat z\hat z}$ contains two terms, one $\O(1)$ and other $\O(f)$. Now $\rho$ can be written as $\rho=\rho_H+\O(f)$ and, as we have stated, $p_{\hat z\hat z}=-\rho+\O(f)$, so that $p_{\hat z\hat z}=-\rho_H+\O(f)$.
On the horizon, where $f(z)\to0$, this has the claimed restricted form (\ref{restricted}).
That is:
\enlargethispage{20pt}
\begin{equation}
[T_{\hat a\hat b}]_H = \left[ \begin{array}{cc|c} 
\rho_H & 0 & 0 \\ 0 &-\rho_H & 0 \\ \hline 0 & 0 & [T_{\hat i\hat j}]_H
\end{array} \right].
\end{equation}
Obviously, as in previous discussions, the remaining $2\times2$ block $[T_{\hat i\hat j}]_H$ can easily be diagonalized.
(A more brutally explicit calculation is carried out in reference~\cite{dirty2}.)
In both sub-cases the on-horizon symmetry implies that the on horizon stress-energy is Hawking--Ellis type I in any static spacetime. 

\subsubsection{Horizons in stationary axisymmetric spacetimes}

To see that this enhanced on-horizon symmetry also works for stationary spacetimes with axial symmetry is very much trickier~\cite{dirty3}.
(The current argument is similar but not quite identical to that in reference~\cite{dirty3}.)
First, note that stationary axial symmetry is enough to block diagonalize both metric and stress-energy:
\begin{equation}
g_{ab} = \left[ \begin{array}{cc|cc} 
* & * & 0 & 0\\ * &*& 0&0\\ \hline 
0 & 0 & *&0 \\ 0 & 0 & 0 & *
\end{array} \right]; \qquad\qquad
T_{ab} = \left[ \begin{array}{cc|cc} 
* & * & 0 & 0\\ * &*& 0&0\\ \hline 
0 & 0 & *&* \\ 0 & 0 & * & *
\end{array} \right].
\end{equation}
To isolate the near-horizon behaviour we set up an ADM-like decomposition
\begin{equation}
g_{ab} = \left[ \begin{array}{cc|cc} 
-f(r,\theta)+ g_{\phi\phi}(r,\theta) \omega(r,\theta)^2 & g_{\phi\phi}(r,\theta) \omega(r,\theta) & 0 & 0\\ g_{\phi\phi}(r,\theta) \omega(r,\theta) & g_{\phi\phi}(r,\theta) & 0&0\\ \hline 
0 & 0 & {h(r,\theta)\over f(r,\theta)} &0 \\ 0 & 0 & 0 & g_{\theta\theta}(r,\theta)
\end{array} \right],
\end{equation}
expressed in the coordinate system $\{t,\phi,r,\theta\}$.
This is carefully constructed to be Lorentzian signature regardless of the \emph{sign} of $f(r,\theta)$. 
This metric has co-tetrad
\begin{equation}
e^{\hat a}{}_{a} = \left[ \begin{array}{cc|cc} 
\sqrt{f(r,\theta)} &0 & 0 & 0\\ 
\sqrt{g_{\phi\phi}(r,\theta)} \omega(r,\theta) & \sqrt{g_{\phi\phi}(r,\theta)} & 0&0\\ \hline 
0 & 0 &\sqrt{{h(r,\theta)\over f(r,\theta)} } &0 \\ 0 & 0 & 0 & \sqrt{g_{\theta\theta}(r,\theta)}
\end{array} \right].
\end{equation}
with corresponding  tetrad
\begin{equation}
e_{\hat a}{}^{a} = \left[ \begin{array}{cc|cc} 
{1\over \sqrt{f(r,\theta)} }&0& 0 & 0\\ 
 -{\omega(r,\theta)\over\sqrt{f(r,\theta)}} &{1\over \sqrt{g_{\phi\phi}(r,\theta)} }& 0&0\\ \hline 
0 & 0 &\sqrt{{f(r,\theta)}\over h(r,\theta)}  &0 \\ 0 & 0 & 0 & {1\over\sqrt{g_{\theta\theta}(r,\theta)}}
\end{array} \right].
\end{equation}

At the horizon, that is when $g_{rr}(r,\theta)\rightarrow\infty$, implying $f(r,\theta)\to0$, we must enforce a constant surface gravity and a constant angular-velocity. 
To accomodate this it is sufficient to demand:
\begin{eqnarray}
\label{E:rigidity}
f(r,\theta) &=& K\times (r-r_H) + \O([r-r_H]^2); \nonumber\\
h(r,\theta) &=&  H + \O(r-r_H); \nonumber\\
\omega(r,\theta) &=& \Omega + \O(r-r_H).
\end{eqnarray}
Here $K$ and $H$ are related to the surface gravity, while $\Omega$ is minus the angular velocity of the horizon.\footnote{Any attempt at making $K$, $H$, or $\Omega$ depend on $\theta$ will result in (naked) curvature singularities at the would-be horizon. Indeed any deviation from the near-horizon behaviour specified in (\ref{E:rigidity}) will result in (naked) curvature singularities at the would-be horizon $r=r_H$.}

Once set up in this manner, a brief but slightly more tedious computation ({\sf Maple} or {\sf Mathematica}) now yields
\begin{equation}
T_{\hat a\hat b} = \left[ \begin{array}{c|c|c|c} 
\rho & \O[(r-r_H)^{1/2}] & 0 & 0\\  \hline
\O[(r-r_H)^{1/2}] &   \O(1) &0 & 0\\ \hline 
0 & 0 &\;-\rho +\O(r-r_H) &\O[(r-r_H)^{1/2}] \\ \hline
0 & 0 & \O[(r-r_H)^{1/2}] &\O(1)
\end{array} \right].
\end{equation}

On the horizon, where $r \to r_H$, this has the claimed restricted form (\ref{restricted}).
Specifically, in the chosen coordinate system the on-horizon stress-energy is actuallly diagonal
\begin{equation}
[T_{\hat a\hat b}]_H = \left[ \begin{array}{c|c|c|c} 
\rho_H & 0 & 0 & 0\\  \hline
0 &   [p_{\hat\phi\hat\phi}]_H &0 & 0\\ \hline 
0 & 0 &\;-\rho_H &0 \\ \hline
0 & 0 & 0 & [p_{\hat\theta\hat\theta}]_H
\end{array} \right].
\end{equation}
(A more brutally explicit calculation is carried out in reference~\cite{dirty3}.)
This stress-energy tensor is now manifestly of Hawking--Ellis type I. 

\subsubsection{Bifurcate Killing horizons}\label{SS:killing}

A bifurcate Killing horizon is composed of two Killing horizons which intersect in the bifurcation $2-$surface.
To see that this enhanced on-horizon symmetry also works for any arbitrary bifurcate Killing horizon, and even more, to any Killing horizon which can be extended to a bifurcate Killing horizon, is rather subtle~\cite{dirty3}. 
The benefit of working with bifurcate Killing horizons is that one does not need to require axisymmetry, the drawback is that the posited existence of the bifurcation 2-surface limits one to eternal black holes, or at the very least, something that is extendable to an eternal black hole. 
(One might also note that if one steps outside of the framework of general relativity, say into the context of the ``analogue spacetimes''~\cite{analogue1,analogue2,analogue3,analogue4}, then there are situations where non-Killing horizons can arise~\cite{non-Killing}.) Be that as it may, let us now see what we can do with bifurcate Killing horizons. 

First, we consider an arbitrary section of a Killing horizon, by definition the vector space perpendicular to this 2-surface can be spanned by 2 null vectors, one of which $K^a$ can be taken to be the Killing vector that is null on the Killing horizon, and the other of which $N^a$ can be taken to be normalized as $K^a N_a = -1$. 
To complete the basis we take 2 orthonormal spacelike vectors $m^a$ and $n^a$ tangent to the 2-surface.
Then
\begin{equation}
g_{ab} = (K_a N_b + N_a K_b) + (m_a m_b + n_a n_b).
\end{equation}
Because we are dealing with a Killing horizon we can without loss of generality choose our basis to be invariant under the Killing flow:
\def\L{{\mathcal{L}}}
\begin{equation}
\L_K K_a = 0; \qquad \L_K\, N_a = 0;  \qquad \L_K\, m_a = 0; \qquad \L_K\, n_a = 0.
\end{equation}

For the stress-energy tensor (indeed for any symmetric tensor) 
\begin{eqnarray}
G_{ab} &=& 
\left\{ G_{KK} K_a K_b + G_{KN} (K_a N_b + N_a K_b) + G_{NN} N_a N_b\right\}\nonumber\\
&&+ 
G_{Km} (K_a m_b+ m_a K_b)  + G_{Kn} (K_a n_b+ n_a K_b) 
\nonumber\\
&&+
 G_{Nm} (N_a m_b+ m_a N_b)  + G_{Nn} (N_a n_b+ n_a N_b) \nonumber\\
&&
+ \left\{ G_{mm} m_a m_b + G_{mn} (m_a n_b + n_a m_b) + G_{nn} n_a n_b\right\}.
\end{eqnarray}
In view of the fact that we are dealing with a Killing horizon $\L_K G_{ab}=0$, so all of the coefficients above are constant along the integral curves of the Killing vector $K^a$.

\enlargethispage{10pt}
Now on any section of the Killing horizon it is a standard result that the Killing vector is a null eigenvector
\begin{equation}
G_{ab} \, K^b = \lambda K_a.
\end{equation}
See for instance reference~\cite[page 333, equation (12.5.22)]{Wald}.\footnote{
What Wald precisely says is that  on any Killing horizon $R_{ab} K^a K^b=0$.
(And the proof is rather roundabout.) 
But since $K^a$ is a null vector on the Killing horizon, this implies $R_{ab} K^b \propto K_a$ on the Killing horizon.
This in turn implies $G_{ab} K^b = \lambda K_a$ on the Killing horizon. 
Note $\lambda$ does not have to be nonzero; in fact in Schwarzschild spacetime it is trivially zero.
}

But
\begin{equation}
G_{ab} K^b = -G_{KN} \, K_a - G_{NN} \,N_a - G_{Nm} \,m_a - G_{Nn} \,n_a.
\end{equation}
Consequently $G_{NN}=G_{Nm}=G_{Nn}=0$, and on any arbitrary section $S$ of the Killing horizon
\begin{eqnarray}
[G_{ab}]_S &=& 
\left\{ G_{KK} K_a K_b + G_{KN} (K_a N_b + N_a K_b)\right\}\nonumber\\
&&+ 
G_{Km} (K_a m_b+ m_a K_b)  + G_{Kn} (K_a n_b+ n_a K_b) 
\nonumber\\
&&
+ \left\{ G_{mm} \,m_a m_b + G_{mn} (m_a n_b + n_a m_b) + G_{nn} \,n_a n_b\right\}
\end{eqnarray}

Now at the bifurcation 2-surface, $N^a$ being perpendicular to the 2-surface, must be parallel to the ``other'' Killing vector that is null on the second sheet of the bifurcate horizon. Thus, \emph{on the bifurcation 2-surface} we must \emph{also} have
\begin{equation}
G_{ab} \, N^b \propto N_a.
\end{equation}
But, explicitly calculating $G_{ab} N^b$ this now implies $G_{KK}=G_{Km}=G_{Kn}=0$, so on the bifurcation 2-surface $B$ we have
\begin{eqnarray}
[G_{ab}]_B &=& 
G_{KN} (K_a N_b + N_a K_b)
+ \left\{ G_{mm} m_a m_b + G_{mn} (m_a n_b + n_a m_b) + G_{nn} n_a n_b\right\}.
\nonumber\\&&
\end{eqnarray}
Once this has been established on the bifurcation 2-surface, the Killing symmetry of 
each individual sheet lets us extend this result to the full Killing horizon.

Thus the on-horizon stress-energy is indeed block diagonal and indeed we can write
\begin{equation}
[G_{ab}]_H = \left[\begin{array}{cc|c} \rho_H &0 & 0 \\ 0 & -\rho_H & 0 \\ \hline 
0 &0&  [G_{mn}]_H \end{array} \right]
\end{equation}
This is manifestly Lorentz diagonalizable, and so manifestly of Hawking--Ellis type~I.

\subsubsection{Summary: On horizon stress-energy}\label{SS:killing}

On any Killing horizon that may be present, be it static, stationary, or bifurcate, the stress-energy tensor is automatically Hawking--Ellis type~I. 

\subsection{On-axis behaviour in stationary (3+1) spacetimes}\label{SS:axisymmetric}

Let us consider a (3+1) stationarity axisymmetric spacetime. 
We now adopt ``quasi-cylindrical'' coordinates $x^a=(t,\phi,r,z)$. Then
\begin{equation}
g_{ab}(r,z) = \left[\begin{array}{cc|cc}
g_{tt} &g_{t\phi} &0&0\\ g_{t\phi} & g_{\phi\phi} &0&0\\
\hline
0 & 0 & g_{rr} & g_{rz} \\ 0 & 0& g_{rz} &g_{zz}
\end{array}\right]
\end{equation}
We can always without loss of generality define the $r$ coordinate by setting $g_{\phi\phi}=r^2$, so that little circles around the axis of rotation have circumference $C=2\pi r$. We  can also without loss of generality choose to set $g_{rz}=0$. 
Then
\begin{equation}
g_{ab}(r,z) = \left[\begin{array}{cc|cc}
g_{tt} &g_{t\phi} &0&0\\ g_{t\phi} & r^2 &0&0\\
\hline
0 & 0 & g_{rr} & 0\\ 0 & 0& 0 &g_{zz}
\end{array}\right]
\end{equation}
The four remaining free functions $g_{tt}(r,z)$, $g_{tr}(r,z)$, $g_{rr}(r,z)$, $g_{zz}(r,z)$, all need to be regular on the rotation axis.

Now we adopt an ADM-like parameterization in terms of ``lapse'' and ``shift'':
\begin{equation}
g_{ab}(r,z) = \left[\begin{array}{cc|cc}
-N(r,z)^2 + r^2 \omega(r,z)^2 &\;\; r^2 \omega(r,z) &0&0\\ r^2 \omega(r,z) & r^2 &0&0\\
\hline
0 & 0 & g_{rr}(r,z)&0\\ 0&0&0& g_{zz}(r,z)
\end{array}\right].
\end{equation}
Then
\begin{equation}
\qquad \det(g) = - N(r,z)^2 \; r^2 \; g_{rr}(r,z)\; g_{zz}(r,z).
\end{equation}

\enlargethispage{10pt}
This is enough to tell us that  (everywhere in the spacetime) 
the non-zero elements of $G_{ab}$ are constrained by 
\begin{equation}
G_{ab} = \left[\begin{array}{cc|cc}
* &*&0&0\\  
{}*& *&0&0\\
\hline
0 & 0 &*&*\\
0 & 0 & * &*
\end{array}\right].
\end{equation}
This is already enough (everywhere in the spacetime) to preclude a stress-energy tensor of type III.

Now (since $g_{rr}$ and $g_{zz}$ are both positive, set $g_{rr}=h^2$ and $g_{zz}=k^2$) we can define a co-tetrad by
\begin{equation}
e^{\hat a}{}_a (r,z) = \left[\begin{array}{cc|cc}
N(r,z)  & 0&0&0\\ 
r  \omega(r,z)  & r&0&0\\
\hline
0 & 0 & h(r,z)&0\\
0 & 0 & 0 & k(r,z)
\end{array}\right],
\end{equation}
and a corresponding tetrad by
\begin{equation}
e_{\hat a}{}^a (r,z) = \left[\begin{array}{cc|cc}
{1\over N(r,z)}  & -{\omega(r,z)\over N(r,z)}&0&0\\ 
0 & {1\over r}&0&0\\
\hline
0 & 0 & {1\over h(r,z)}&0\\
0 & 0 &0&  {1\over k(r,z)}
\end{array}\right].
\end{equation}
Then in the corresponding orthonormal basis we have
\begin{equation}
G_{\hat a \hat b}(r,z) =  e_{\hat a}{}^a(r,z) \; e_{\hat b}{}^b(r,z)\; G_{ab}(r,z) = 
\left[\begin{array}{cc|cc}
*&* &0&0\\ 
{}* &* &0&0\\
\hline
0 & 0 &*&*\\
0 & 0 &*&*\\
\end{array}\right].
\end{equation}

What can we say on-axis? (That is, $r=0$.)
To avoid a conical singularity at $r=0$, we need $g_{rr}(r,z) = 1 + o(r)$, so $h(r,z) = 1 + o(r)$. 
Looking along the rotation axis at $r=0$ we can also without loss of generality set $g_{zz}(r,z) = 1 + o(r)$, so $k(r,z) = 1 + o(r)$. 
(That is, $z$ is chosen to be proper distance along the axis of rotation.)
To be more precise, demanding regularity near the rotation axis forces us to set
\begin{equation}
N(r,z) = N_0(z) + {1\over2} N_2(z) r^2 + ...; \qquad
\omega(r,z) = \omega_0(z) + {1\over2} \omega_2(z) r^2 + ...; \qquad
\end{equation}
\begin{equation}
h(r,z) =1 + {1\over2} h_2(z) r^2 + ....; \qquad k(r,z) =1 + {1\over2} k_2(z) r^2 + ....; 
\end{equation}
A brief calculation ({\sf Maple} or {\sf Mathematica}) shows that at  the axis of rotation ($r=0$) one has:
\begin{equation}
(G_{\hat a \hat b})_{r=0} =  (e_A{}^a \; e_B{}^b\; G_{ab})_{r=0} =
\left[\begin{array}{cc|cc}
(G_{\hat t \hat t})_0 &0 &0&0\\ 
  0 &(G_{\hat \phi \hat \phi})_0 &0\\
\hline
0 & 0 &(G_{\hat r \hat r})_0 &0\\
0 & 0 & 0 & (G_{\hat z \hat z})_0
\end{array}\right].
\end{equation}
Here we explicitly have:
\begin{equation}
(G_{\hat t \hat t})_0 = h_2(z)-2 k_2(z);
\end{equation}
\begin{equation}
(G_{\hat \phi \hat \phi})_0 = (G_{\hat r \hat r})_0 = 
k_2(z) + {N_2(z)\over N_0(z)}  + {[N_0(z)]_{,zz}\over N_0(z)};
\end{equation}
\begin{equation}
(G_{\hat z \hat z})_0 =  {2 N_2(z)\over N_0(z)} - h_2(z);
\end{equation}
with all other stress-energy components being zero. 
This is manifestly diagonal, and so manifestly Hawking--Ellis type~I.
Note that the use of the tetrad/co-tetrad formalism and orthonormal basis is essential to getting this to work cleanly. 

In short: The stress-energy is always Hawking--Ellis type~I on the axis of rotation of any axisymmetric stationary (3+1) spacetime.

\subsection{Synchronous cosmological spacetimes}\label{SS:synchronous}

Instead of focusing on spacetimes with timelike Killing vectors, we shall now explore cosmological spacetimes with a (3+1) metric that is block diagonalizable into $(1\times1)\oplus(3\times3)$ blocks. We shall further assume that in this spacetime $g_{00}=-1$ and, therefore, we have a spacetime in so-called ``synchronous form'':
\begin{equation}\label{synch}
\d s^2=-\d t^2+g_{ij}(t,x^k)\,\d x^i\d x^j.
\end{equation}
Here the 3-metric $g_{ij}$ has Euclidean signature and, for the moment, the functions $g_{ij}(t,x^k)$ can depend on any or all coordinates. 
The unit timelike vector orthogonal to the spatial slicing is $n^a=(1,0,0,0)$. So, we have $n_a=(-1,0,0,0)$ and
\begin{equation}
K_{ij}=n_{i;j}=\Gamma^0{}_{ij}=\frac{1}{2}\,\dot g_{ij}.
\end{equation}
Condition (\ref{condition}), that is
\begin{equation}
K^{ij}{}_{:j} -  g^{ij} K_{:j} =0,
\label{condition2}
\end{equation}
can now be analyzed in a simpler way under these more restrictive assumptions. Any synchronous cosmology  (\ref{synch})  satisfying (\ref{condition2}) will have a stress-energy tensor of the form 
$R_{ti}=0=G_{ti}$. So, as we have already discussed, using ordinary spatial 3-rotations it can be seen that $T_{ab}$ would then be manifestly of Hawking--Ellis type~I. 

\enlargethispage{40pt}
The simplest (trivial) case in which this condition is satisfied is a subclass of the static spacetimes, that have already been analyzed above. In this case, we have 
\begin{equation}
\dot g_{ij}=0\qquad\Longrightarrow\qquad K_{ij}=0,
\end{equation}
and the condition is clearly satisfied.

\subsubsection{Bianchi type I spacetimes}\label{SSS:B_I}

The next simplest case that we can consider of a block diagonalizable metric of the form (\ref{synch}) with $\dot g_{ij}\neq0$ is one for which the spatial slicing is proportional to a flat Euclidean 3-space: $g_{ij}(t,x^k) \to h_{ij}(t)$. In this case $g_{ij,k}=0$ and
we can write the metric as
\begin{equation}\label{B-1}
\d s^2=-\d t^2+h_{ij}(t)\,\d x^i\d x^j.
\end{equation}
This is simply the  \emph{Bianchi~I cosmology}, the simplest of the Bianchi homogeneous spacetimes~\cite{Ryan-Shepley}.
Since
\begin{equation}
K_{ij}=\frac{1}{2}\,\dot h_{ij}(t),
\end{equation}
is now position independent, 
we have $K_{:j}=0$. Then the sufficient condition for a guaranteed type~I stress-energy tensor is just
\begin{equation}
K^{ij}{}_{:j} =0.
\end{equation}
Since the connection in the spatial 3-slices vanishes, this equation is clearly satisfied.
Hence all \emph{Bianchi~I spacetimes} are Hawking--Ellis type I.
It is worth noting that spatially-flat FLRW spacetimes belong to this class. 

\subsubsection{Self-similar spatial slicings}\label{SSS:self}

\enlargethispage{10pt}
We now allow curvature for the spatial slicings, but impose self-similarity so that the spatial slices have the same shape but can differ in overall scale factor. That is, we set   $g_{ij}(t,x) \to a(t)^2 \; h_{ij}(x)$ so that
\begin{equation}
\d s^2=-\d t^2+a(t)^2\,h_{ij}(x^k)\,\d x^i\d x^j.
\end{equation}
Here, for the moment, we allow the $h_{ij}(x^k)$ to be generic functions of the spatial coordinates.
Geometries of this type are still much more general than FLRW, allowing for both spatial inhomogeneities and anisotropies.  
In this situation, we have 
\begin{equation}
K_{ij}=a\dot a \, h_{ij},\qquad {\rm and} \qquad K=3\,\frac{\dot a }{a^3}.
\end{equation}
Under these conditions the sufficient condition for a type~I stress-energy tensor can be written as
\begin{equation}
h^{ij}{}_{:j} =0.
\end{equation}
But is trivially satisfied for any 3-metric $h^{ij}(x^k)$, and, therefore the stress-energy tensor is Hawking-Ellis type I.

\subsubsection{Single-mode restriction of Bianchi types II to IX }\label{SSS:B-2-to-9}

General Bianchi cosmologies (the spatially homogeneous cosmologies) can all be written in the form
\begin{equation}
\d s^2 = - \d t^2 + h_{IJ}(t) \, \omega^I \omega^J.
\end{equation}
Here the $\omega^I$ are the 1-forms dual to the invariant basis vectors used in setting up the Bianchi classification of homogeneous 3-geometries. See reference~\cite{Ryan-Shepley} for an extensive discussion. (Note especially table 6.1 on pages 110--113.)
Since $h_{IJ}(t)$ is a symmetric $3\times3$ matrix, the general Bianchi cosmologies depend on up to 6 interacting ``modes'', six interacting free functions $h_{IJ}(t)$.
Suppose now that we restrict attention to a single mode by setting $ h_{IJ}(t) \to a(t)^2\; h_{IJ}$,
where the $h_{IJ}$ are now constants. Then
\begin{equation}
ds^2 = - dt^2 + a(t)^2 \left\{ h_{IJ} \, \omega^I \omega^J\right\} =
 - dt^2 + a(t)^2 \left\{ h_{ij}(x) \, dx^i dx^j\right\}. 
\end{equation}
That is, single-mode Bianchi cosmologies of this form are automatically spatially self-similar, and the discussion above applies.
Consequently the stress-energy tensor is Hawking--Ellis type I.

It is easy to note that all three \emph{Friedmann--Lema\^itre--Robertson--Walker spacetimes} are of this form.
As expected all three FLRW spacetimes (which are isotropic spatially self-similar sub-cases of  Bianchi types~I, V, and IX respectively) have a stress-energy of Hawking--Ellis type~I. 

Unfortunately we can give no really general arguments for other more general multi-mode Bianchi cosmologies, and at best one has to resort to case-by-case analyses. 

\section{Conclusions}\label{S:conclusions}

We have explicitly demonstrated that, when considering self-consistent solutions of the Einstein equations, the presence of symmetry often severely restricts the nature of the stress-energy tensor under the Hawking--Ellis  (Segr\'e--Pleba\'nski)  classification. 
The same considerations also apply to any proposed modifications or extensions of Einstein gravity which have equations of motion that 
can be rearranged into the form $G_{ab} = 8 \pi [T_\mathrm{effective}]_{ab}$.
Then at a minimum our results constrain the Hawking--Ellis classification of  $[T_\mathrm{effective}]_{ab}$.

Working in the test-field limit it is rather easy to find examples of Hawking--Ellis types I and IV. (Hawking--Ellis types II and III are perturbatively unstable; either under numerical round-off error or under generic physical perturbations.) 

Working within the framework of self-consistent solutions of the Einstein equations, type IV is often excluded. Indeed in (3+1) dimensions the stress-energy tensor is guaranteed to be Hawking--Ellis type I in at least the following situations: 
\vspace{-10pt}
\begin{itemize}
\itemsep-3pt
\item  In the domain of outer communication of any static spacetime.
\item On any Killing horizon (static, stationary, bifurcate).
\item On the axis of rotation of any axisymmetric stationary spacetime.
\item In any Bianchi type I cosmology.
\item In any single-mode restriction of the Bianchi type II to type IX cosmologies.
\end{itemize}
\vspace{-10pt}
This list is not necessarily exhaustive, and we are actively seeking further examples of this or similar behaviour. 

It is worthwhile to note that if it could be proved that the stress-energy tensor of self-consistent geometries has to be type~I in most physically relevant situations, then the study of some implications of semi-classical effects could be simplified. For example, one could study the properties of (possibly regular) black hole spacetimes taking into account Hawking radiation.
In the appendices below we also comment on circular symmetry in (2+1) dimensions, and on dilaton gravity in (1+1) dimensions.

\appendix

\section{Appendix: Circular symmetry in (2+1)-dimensions}\label{S:2+1-D}

For completeness, let us consider circular symmetry in (2+1)-dimensions. 
(Note that (2+1) dimensions is often surprisingly subtle. For instance there \emph{is} a Birkhoff theorem for rotating ``stars'' in (2+1) dimensions~\cite{2+1-a,2+1-b}, so it is well worth the effort to check (2+1)-dimensional physics explicitly.)
Ordering the coordinates as $(t,\phi,r)$,  we can write
\begin{equation}
g_{ab}(r) = \left[\begin{array}{cc|c}
g_{tt} &g_{t\phi} &0\\ g_{t\phi} & g_{\phi\phi} &0\\
\hline
0 & 0 & g_{rr}  
\end{array}\right],
\end{equation}
where the metric components only depend on $r$.
Without loss of generality we define the $r$ coordinate by setting $g_{\phi\phi}=r^2$, 
then
\begin{equation}
g_{ab}(r) = \left[\begin{array}{cc|c}
g_{tt} &g_{t\phi} &0\\ g_{t\phi} & r^2 &0\\
\hline
0 & 0 & g_{rr}
\end{array}\right].
\end{equation}
We now adopt an ADM-like decomposition, then:
\begin{equation}
g_{ab}=\left[ \begin{array}{cc|c}
-N(r)^2 + r^2 \omega(r)^2 & r^2 \omega(r)&0\\
r^2 \omega(r) & r^2 & 0\\
\hline
0&0&h(r)^2
\end{array}\right]_{ab}.
\end{equation}
Thence
\begin{equation}
g = \det(g_{ab}) = - N(r)^2 r^2 h(r)^2.
\end{equation}
Now we define a co-triad by
\begin{equation}
e^A{}_a (r) = \left[\begin{array}{cc|c}
N(r)  & 0&0\\ 
r  \omega(r)  & r&0\\
\hline
0 & 0 & h(r)
\end{array}\right],
\end{equation}
and the corresponding triad by
\begin{equation}
e_A{}^a (r) = \left[\begin{array}{cc|c}
{1\over N(r)}  &  0 &0\\ 
 -{\omega(r)\over N(r)}   & {1\over r}&0\\
\hline
0 & 0 & {1\over h(r)}
\end{array}\right].
\end{equation}

What can we say at the centre of rotation? (That is, $r=0$.)
To avoid a conical singularity at $r=0$, at a minimum we need $h(r) = 1 + o(r)$. 
To be more precise, demanding regularity near the centre of rotation forces us to set
\begin{equation}
N(r) = N_0 + {1\over2} N_2 \,r^2 + ...; \qquad
\omega(r) = \omega_0 + {1\over2} \omega_2\, r^2 + ...; \qquad
\end{equation}
\begin{equation}
h(r) =1 + {1\over2} h_2\, r^2 + ....
\end{equation}
with $N_0 > 0$ so that the centre has finite redshift. In contrast $N_2$, $\omega_0$,  $\omega_2$, and $h_2$ are typically though not necessarily non-zero.

A brief calculation ({\sf Maple} or {\sf Mathematica}) shows that at  the axis of rotation ($r=0$) one has:
\begin{equation}
(G_{AB})_{r=0} =  (e_A{}^a \; e_B{}^b\; G_{ab})_{r=0} =
\left[\begin{array}{cc|c}
(G_{\hat t \hat t})_0 &0 &0\\ 
  0 &(G_{\hat \phi \hat \phi})_0 \\
\hline
0 & 0 &(G_{\hat r \hat r})_0 
\end{array}\right].
\end{equation}
Here we explicitly have:
\begin{equation}
(G_{\hat t \hat t})_0 = h_2;
\qquad\qquad
(G_{\hat \phi \hat \phi})_0 = (G_{\hat r \hat r})_0 = 
{N_2\over N_0};
\end{equation}
with all other stress-energy components being zero.
This is manifestly diagonal, and so manifestly type I.
Note that the use of the triad/co-triad formalism and orthonormal basis is essential to getting this to work cleanly. 

In short: The stress-energy is always Hawking--Ellis type~I at the centre of rotation of any circularly symmetric stationary (2+1) spacetime. 

\section{Appendix: (1+1)-dimensions}\label{S:2+1-D}

For completeness, let us finally consider the situation in (1+1)-dimensions. 
(Note that (1+1) dimensions one can often obtain exact results~\cite{Visser:1996-III},  so it is well worth the effort to check (1+1)-dimensional physics explicitly.) Note that Hawking--Ellis type~III is automatically excluded in (1+1) dimensions, though all three of Hawking--Ellis types~I, II, and IV are \emph{a priori} possible. Furthermore in (1+1) any nonzero vector field is hypersurface orthogonal, so there is no distinction between static and stationary.

\subsection{Test field stress-energy}

Let us first consider a test-field computation: we consider a massless minimally coupled scalar field and follow the analysis of~\cite{Visser:1996-III}, which is in turn based on discussion in~\cite{Fulling,B&D}.
We take the orthonormal form of the stress-energy to be
\begin{equation}
T_{\hat a\hat b} = \left[\begin{array}{cc} 
\rho & f \\ f &p 
\end{array}\right]
\end{equation}
Then the two Lorentz invariant eigenvalues are
\begin{equation}
\lambda =\frac{1}{2}(p-\rho) \pm\frac{1}{2} \sqrt{(\rho+p)^2-4 f^2}.
\end{equation}
In the Boulware of Hartle--Hawking vacuum states the flux is zero, so the stress-energy is automatically type~I. So let us focus on the Unruh vacuum state. Specifically, for the (1+1) dimensional version of Schwarzschild 
\begin{equation}
\d s^2 = -(1-2m/r) \d t^2 + {\d r^2\over1-2m/r},
\end{equation}
defining $z=2m/r$, the stress-energy in the Unruh vacuum is easily computed to be~\cite{Visser:1996-III}
\begin{eqnarray}
\rho &=& p_\infty \;{1-16z^2+14z^4\over2(1-z)};\nonumber\\
p &=& p_\infty \;{1-2z^4\over2(1-z)};\nonumber\\
f &=& p_\infty \; {1\over2(1-z)}.
\end{eqnarray}
Consequently
\begin{equation}
\Gamma = (\rho+p)^2 - 4 f^2 =   p_\infty^2\;{4 z^2 (1+z)(1-3z^2)(4-3z^2)\over 1-z}.
\end{equation}
This changes sign at $z\in\left\{ {1\over\sqrt{3}}, 1, {2\over\sqrt{3}} \right\}$. 
The test-field stress-energy is Hawking--Ellis type I for $z\in \left[{1\over\sqrt{3}}, 1\right)$ and $z\geq{2\over\sqrt3}$, and is type IV for $z\in \left[0,{1\over\sqrt{3}}\right)$ and $z\in \left(1,{2\over\sqrt{3}}\right)$.
So test field stress-energies can easily be type IV in (1+1) dimensions.

\subsection{Dilaton gravity}

To go beyond the test field limit one has to choose a specific theory of (1+1) gravity.
The Einstein tensor is identically zero, so the usual Einstein equations are not relevant. The Ricci tensor identically satisfies $R_{ab}={1\over 2} R\, g_{ab}$ and is automatically always type I. To get anything interesting one needs something more subtle such as a dilaton gravity. Let us choose for instance:
\begin{equation}
S = \int \d^2 x \sqrt{-g} \{\varphi R  -U(\varphi)\} + S_\mathrm{matter}.
\end{equation}
Then one has
\begin{equation}
T_{ab} = 2 \nabla_a \nabla_b \varphi- 2g_{ab} \nabla^2 \varphi -g_{ab} U(\varphi),
\end{equation}
Thus, up to a trivial shift, the Lorentz invariant eigenvalues of the stress-energy are determined by the 
Lorentz invariant eigenvalues of the traceless tensor
\begin{equation}
\hat T_{ab} =2 \left( \nabla_a \nabla_b \varphi
- {1\over2}  g_{ab} \nabla^2 \varphi\right).
\end{equation}

Now in (1+1) dimensions any Killing vector must satisfy 
\begin{equation}
\nabla_a K_b \propto \epsilon_{ab}.    
\end{equation}
Furthermore, if there is a Killing vector $K^a$ under which the scalar field is invariant ($K^a\nabla_a\varphi=0$) then in (1+1) dimensions both $K^a$ and $\epsilon^{ab}  \nabla_b \varphi$ are perpendicular to $\nabla_a\varphi$, so they must be parallel to each other:
\begin{equation}
K_a  \propto \epsilon_{ab}  \nabla^b \varphi.
\end{equation}
But then
\begin{equation}
K^a (\nabla_a \nabla_b \varphi) =  K^a (\nabla_b \nabla_a \varphi)    
=  \nabla_b (K^a \nabla_a \varphi)   - (\nabla_b K^a) \nabla_a \varphi
= - (\nabla_b K_a) \nabla^a \varphi.
\end{equation}
This implies
\begin{equation}
K^a (\nabla_a \nabla_b \varphi) \propto (\epsilon_{ba}) \nabla^a \varphi \propto K_b.
\end{equation}
Consequently in (1+1) dimensions any Killing vector is an eigenvector of the dilaton stress-energy tensor. 
If the Killing vector is timelike, then we are done (the stress-energy  is Hawking--Ellis type I).
If the Killing vector is spacelike, then with one spacelike eigenvector, the other must be timelike, then we are again done (the stress-energy  is Hawking--Ellis type I).
The exceptional case is when the Killing vector is null, then the stress-energy is Hawking--Ellis type II.
Therefore for any non-null Killing vector, in (1+1) dimensions the 
stress-energy tensor in dilaton gravity is Hawking--Ellis type I. \footnote{Indeed in any arbitrarily complicated scalar-tensor theory of (1+1) gravity we will have \\ \null \qquad $T_{ab} = A \; \nabla_a \nabla_b \varphi + B \; \nabla_a\varphi \, \nabla_b\varphi + C \; g_{ab}$, and the same argument will apply.}

\section*{Acknowledgments}
PMM acknowledges financial support from the project PID2019-107394GB-I00 (MINECO, Spain). MV acknowledges financial support via the Marsden Fund administered by the Royal Society of New Zealand.

\clearpage
\addcontentsline{toc}{section}{\null\quad\; References}

\end{document}